# Analysis of chemical pathways for n-dodecane/air turbulent premixed flames[1]

D. Dasgupta[1], W. Sun[1], M. Day[2], A.J. Aspden[3], T. Lieuwen[1]

[1]School of Aerospace Engineering, Georgia Institute of Technology, Atlanta, GA, 30332, USA

[2]Center for Computational Sciences and Engineering, Lawrence Berkeley National Laboratory, Berkeley, CA, 94720, USA

[3]School of Engineering, Newcastle University NE1 7RU, UK

Corresponding author: Debolina Dasgupta

School of Aerospace Engineering

270 Ferst Drive

Montgomery Knight Building 0150

Atlanta, Georgia 30332-0150

Email: ddasgupta3@gatech.edu

Phone: 404-490-7868

---

**Abstract**

This paper analyzes turbulence-chemistry interactions for an *n*-dodecane-air flame, focusing on the degree to which fuel oxidation pathways change in turbulent flames relative to their corresponding laminar flames. This work is based on a lean ($\phi$=0.7) *n*-dodecane-air flame DNS database from Aspden et al. (Proc. Combust. Institute, 36 (2017) 2005-2016). The relative roles of dominant reactions that release heat and produce/consume radicals are examined at various turbulence intensities and compared with stretched flame calculations from counterflow flames and perfectly stirred reactors. These results show that spatially integrated (i.e. integrated heat release or radical production rate metrics averaged over the entire flame) chemical pathways are relatively insensitive to turbulence intensity and mimic the behavior of stretched flames. In other words, the contribution of a given reaction to heat release or radical production, integrated over the entire flame, is nearly independent to turbulence. Localized analysis conditioned on topological feature of the flame and on temperature is also performed. The former analysis reveals that much larger alteration of pathways occurs in the positively-curved regions of the flame. The latter localized analysis shows that peak activity in the low temperature (i.e. below 1200K) region shift towards *higher* temperatures with increases in Karlovitz number. This result is particularly interesting given that prior work with lighter fuels showed the opposite behavior suggesting a disparate response of the reactions involved in the fuel oxidation process to increased turbulence.

# 1. Introduction

Turbulence has well-known impacts on the large scale topology and wrinkling of a premixed flame, and may modify the internal structure of the flame through processes associated with local, time-dependent strain, curvature, and convective transport. The influence of turbulence on the chemical reaction pathways is less well understood; this question is particularly significant given that kinetic mechanisms are generally validated and benchmarked with measured data from canonical laminar flames, such as bomb reactors or steady laminar flames. In addition, flamelet modeling approaches generally use libraries developed from laminar calculations for unstretched and stretched flames (for example, [1]) and it is essential to understand their validity in describing chemistry in highly turbulent flow fields.

The objective of this paper is to consider the extent to which chemical pathways vary with turbulence levels. A detailed discussion of the potential mechanisms through which turbulence could alter reaction pathways is given in our previous study [2], and is briefly summarized here. First, unsteady stretching and internal mixing in the flame by turbulent disturbances can alter the correlations between species concentrations and temperatures and hence reaction rates. Similarly, the relative contributions of different reactions could be altered in turbulent flames due to unsteady kinetic and diffusive effects, such as if one species can rapidly adjust its local concentration to fluctuating local conditions, while another cannot. This also leads to differences in relative values of concentrations of different species at a given location or isotherm. These can, in turn, cause local changes to the reaction rates that may disturb the chemical pathways. Finally, small turbulence length scales can lead to convective stirring within the flame itself, and give rise to spatially varying convective transport of species within the reaction volume.

The effect of turbulence on the flame structure and chemical pathways has been explored for a number of fuels, including hydrogen, methane, propane, *n*-heptane and *n*-dodecane [3-7]. Generally, it is found that global analysis (statistics averaged over the entire reaction volume) of chemical pathways exhibit limited changes, in comparison to the laminar flames counterpart. For example, Lapointe et al.[7] compared the dominant three fuel consumption reactions for *n*-heptane and observed ~1-3% change between the relative role of each reaction toward the overall fuel consumption rate between turbulent and laminar cases. They also compared two high heat-release reactions: $CH_3+O \rightarrow CH_2O+H$ and $HCO+H \rightarrow CO+H_2$, and noted less than 1% change in their contributions to the total heat release. Additionally, LTC studies for n-heptane flames by Savard et al.[8] suggest fuel oxidation for cool flames occur through the same chemical pathways in laminar and turbulent flames. Similarly, studies of $H_2$/air [2, 6] and $CH_4$/air [9] flames have shown relatively minor changes in the spatially integrated dominant chemical pathways. For example, the contribution of $H+O_2(+M) \rightarrow HO_2(+M)$ (the primary heat releasing reaction) to the total heat release for $H_2$/air flames changed by 20% between unstretched laminar flame and $Ka$=36 turbulent flames and changed by only 5% between $Ka$=1 to $Ka$=36 for the turbulent flames. However, much larger changes in these contributions were noted for reactions of secondary influence. For example, $H+OH+M \rightarrow H_2O+M$ (the third largest heat release contributor) is the most sensitive to increasing turbulence, and its contribution to heat release roughly doubles as Karlovitz number increases from 1 to 36. Overall, relatively minor changes in chemical pathways occur for the dominant contributors to heat release or fuel consumption, while relatively significant changes can occur for secondary contributors.

Much larger changes in reaction pathways are observed locally in the flame, suggesting more nuances to the question of how chemical pathways are altered by turbulence. A nearly universal observation from temperature-conditioned analysis is an increase in reactivity at low temperatures with increasing $Ka$, likely due to convective stirring within the flame by small scale eddies. For example, DNS of lean premixed $H_2$ flames [4, 5] showed an increase in low temperature heat release (~2-3 times the laminar heat release) due to the reactions $H+O_2(+M) \rightarrow HO_2(+M)$, $HO_2+H \rightarrow OH+OH$ and $HO_2+OH \rightarrow H_2O+O_2$ and a decorrelation in fuel consumption and heat release in regions of strong negative curvature. This is primarily attributed to an increased radical pool at lower temperatures. Similarly, there is an increase in low temperature heat release in lean, premixed methane flames due to the reactions $H+O_2(+M) \rightarrow HO_2(+M)$ and $H+CH_2O \rightarrow HCO+H_2$ [5].

Another notable observation is the broadening of the preheat zone for a number of fuels. Aspden et al. noted thermal preheat zone thickening by a factor of 1.5-2.5 from $Ka$=1 to $Ka$=36 for methane flames [10] and 2.5-4 from $Ka$=1 to $Ka$=36 for n-dodecane flames[11]. Similarly, Savard et al.[12] observed a factor of ten increase in preheat zone thickness compared to its laminar flame counterpart for n-heptane flames at high Karlovitz numbers.

The above examples show turbulence effects on chemical pathways and flame structure using temperature as the variable for locally conditioning the analysis. Another way to consider local effects is to consider turbulence effects on specific geometric topological regions of the flame. For $CH_4$/air flames, the reaction pathways are a function of flame curvature [9]. For example, the dominant heat release reaction shifted from $O+CH_3 \rightarrow H+CH_2O$ to $OH+CO \rightarrow CO_2+H$ in the positively-curved elements relative to the overall flame. On the other hand, the negatively-curved elements and saddle-point exhibited no such shifts[9]. Day et al. [3] examined 2D DNS of $H_2$-$CH_4$-air turbulent flames and noted three different regions of the flame front: (a) intense burning regions with positive curvature, (b) weak burning regions with negative curvature (c) large scale flame folding regions (regions where $H_2$ consumption is negligible but $C_2$ hydrocarbon concentrations are high). They investigated corresponding changes in $C_1$ and $C_2$ kinetic pathways for each of these regions. The $C_1$ pathway via $CH_3O$ varied by a factor of three over these regions, increasing from 4% in region (a) to 8% in region (b) to 12% in region (c). Similarly, the $C_2$ pathway shifted from 3% to 4% to 5% in regions (a), (b) and (c) respectively. The study suggests that a relative strengthening of minor pathways could occur locally on the flame surface.

This paper is a continuation of the effort to understand the global and local changes in chemical pathways for different fuels. The work presented here focuses on lean premixed *n*-dodecane/air flames, which have the property that the deficient reactant is heavier than oxygen and has a *Le*>1. Specifically, we are interested in how the progress rates of various key reactions, and their heat release, are affected by turbulence. Section (A) of the results provides an analysis for integrated quantities averaged over the entire flame surface. Section (B) discusses results conditioned on geometric topology. Section (C) presents the local effects of turbulence conditioned on temperature.

## 2. Methodology

We utilize the DNS data set from Aspden et al.[11], which consists of a nominally one-dimensional, lean ($\phi$=0.7), premixed turbulent *n*-dodecane/air flame. The DNS is based on a Low-Mach number reacting flow model with mixture-averaged transport for molecular diffusion, Soret and Dufour transport, gravity and radiative processes are neglected[13, 14]. The You et al. [15] model for reaction kinetics, thermodynamic properties and transport coefficients is used. This model consists of 56 chemical species and 289 reaction steps. Appendix A discusses the sensitivity of this model to different metrics for various model reactors/flames, including heat release, and the production rates of key species, and compares to those exhibited by other models of Luo et al. [16] and Narayanswamy et al. [17]. It is concluded that much of the analysis to follow is rather insensitive to the choice of detailed model.

The DNS utilizes reactant temperature $T^u$=298 K at a uniform pressure of 1 atm. A high aspect ratio domain (8:1:1) is used for the simulations with lateral periodic boundary conditions, free-slip fixed wall at one end and outflow at the other. A time-dependent, density-weighted forcing term in the momentum equations throughout the volume maintains the turbulence with an integral length scale of $l/l_O = 1$, where $l$ and $l_O$ denote the integral length scale and unstrained laminar flame length respectively. A flame is initialized near the top of the domain, and propagates downwards through the turbulent fuel mixture until it reaches the bottom of the domain. During this evolution, the flame reaches a quasi-stationary burning rate from which flame statistics are extracted for analysis here. The DNS averages and conditional means are calculated from five time steps, each spaced one integral time scale apart. Five cases of increasing turbulence intensity (Karlovitz number = 1, 4, 12, 36 and 108) are analyzed where Karlovitz number is defined as:

$$Ka_F^2 = \frac{u'^3}{s_{L_O}^3} \frac{l_O}{l} \qquad (1)$$

Here $u'$ is the turbulent rms velocity and $s_{L_O}$ is the unstretched laminar flame speed. For these cases, $s_{L_O} \approx 22.6$ cm/s and $l_O \approx 520$ microns[11]. Table 1 summarizes these cases.

**Table 1: Summary of cases [11]**

| Case | $u'$ (m/s) | $u'/s_{L_0}$ | $Ka_F$ |
|---|---|---|---|
| A | 0.226 | 1.00 | 1 |
| B | 0.570 | 2.52 | 4 |
| C | 1.18 | 5.25 | 12 |
| D | 2.47 | 17.3 | 36 |
| E | 5.12 | 22.7 | 108 |

Following closely the analysis procedures discussed in Dasgupta et al. [1], the "local" flame behavior is quantified via integrated flame sub-volumes, constructed to pass through localized sections of a triangulated isosurface that represents an instantaneous snapshot of the flame following the methodology in Ref. [18]. The volumes extend through the flame from cold fuel to hot products, and have lateral boundaries aligned with integral curves in flame progress. A reference isosurface will be used to associate with each sub-volume the values of local curvature, stretch, and the flame normal. Here, the reference isosurface is taken as an isotherm at the temperature of peak heat release in a steady, unstretched flame at the same equivalence ratio ($\phi$=0.7), $T_{ref}$ = 1460 K. Figure 1 shows the isotherm, $T_{ref}$ = 1460 K (colored by local values of the heat release) for the cases considered. Since the volumetric statistics are integrated normal to the flame, they are generally insensitive to the precise isotherm used. Details of the integration procedure and the subsequent averaging are detailed in Dasgupta et al. [2].

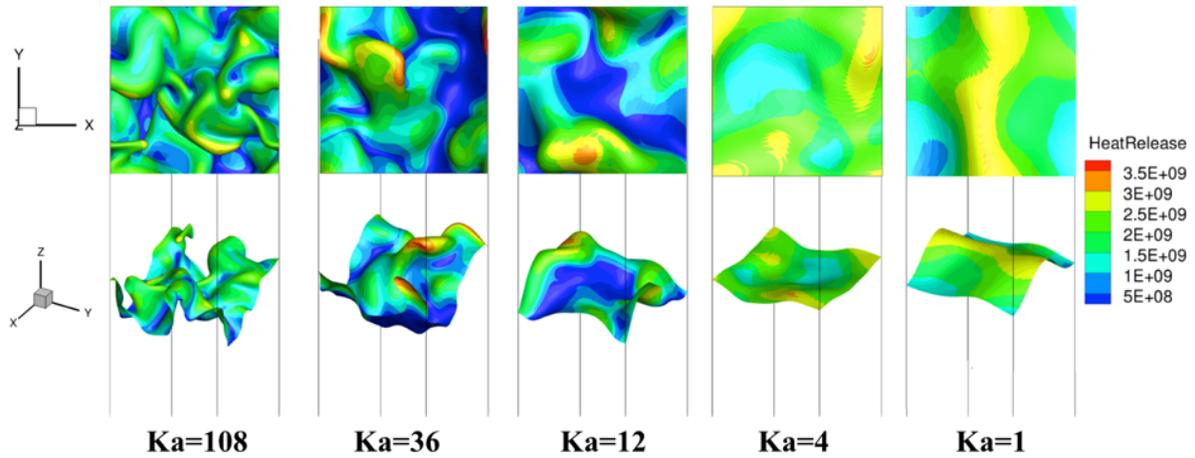

**Figure 1. Isotherm, $T_{ref}$ = 1460 K(colored by heat release). (Top) X-Y slices of the flame surface (bottom) 3D view of the flame surface. The rectangular domain is shown in black lines.**

A corresponding procedure was performed on unstretched premixed flame (using PREMIX[19]), a stretched, premixed laminar flames (using OPPDIF[20]) and perfectly stirred reactor (using PSR[21]) to compare the turbulent flame-chemistry interaction with simple laminar models. A typical maximum number of 6000 grid points was used to ensure convergence of individual 1D OPPDIF runs with a value of 0.01 for adaptive grid control based on profile curvature and gradient (CURV and GRAD). The radial velocity is set to zero at the inlets. The stretch rate, κ, is given by the maximum value of $-du/dx$ between the inlet and the first minima in the axial velocity profile. The extinction stretch rate, $\kappa_{ext}$ is 245 1/s. Figure 2(a) plots the flame speed and temperature dependence upon the normalized stretch rate. The $\kappa$=0 result comes from PREMIX. The PSR is set up with the same inlet fuel/air composition and inlet temperature as DNS. It was run with different residence times, $\tau_{res}$, down to values approaching extinction, given by $\tau_{ext}$=0.16 ms. Figure 2(b) plots the temperature variation as a function of residence time. For reference, Figure 2(c) plots the temperature variation for the DNS flames with Karlovitz number. The flame normal is defined using temperature gradients[18]. All data is interpolated on these normals. The average maximum temperature is obtained by taking the mean of the maximum temperature along all flame normals. The error bars indicate one standard deviation from the mean value of this maximum temperature at every *Ka*. This mean changes by ~80K from *Ka*=1 to *Ka*=108.

Comparing the temperature changes for the three calculations, there is a roughly 300K, 200K, and 80K difference for the PSR, stretched flame, and DNS. In other words, the PSR temperature changes the most and the DNS the least – we will return to this point later as it appears to be a key driver behind the much larger changes in chemical pathways for the PSR with residence time than is observed for the stretched flames, as well as a possible reason for the near insensitivity of integrated reaction metrics to Karlovitz number for the turbulent flame.

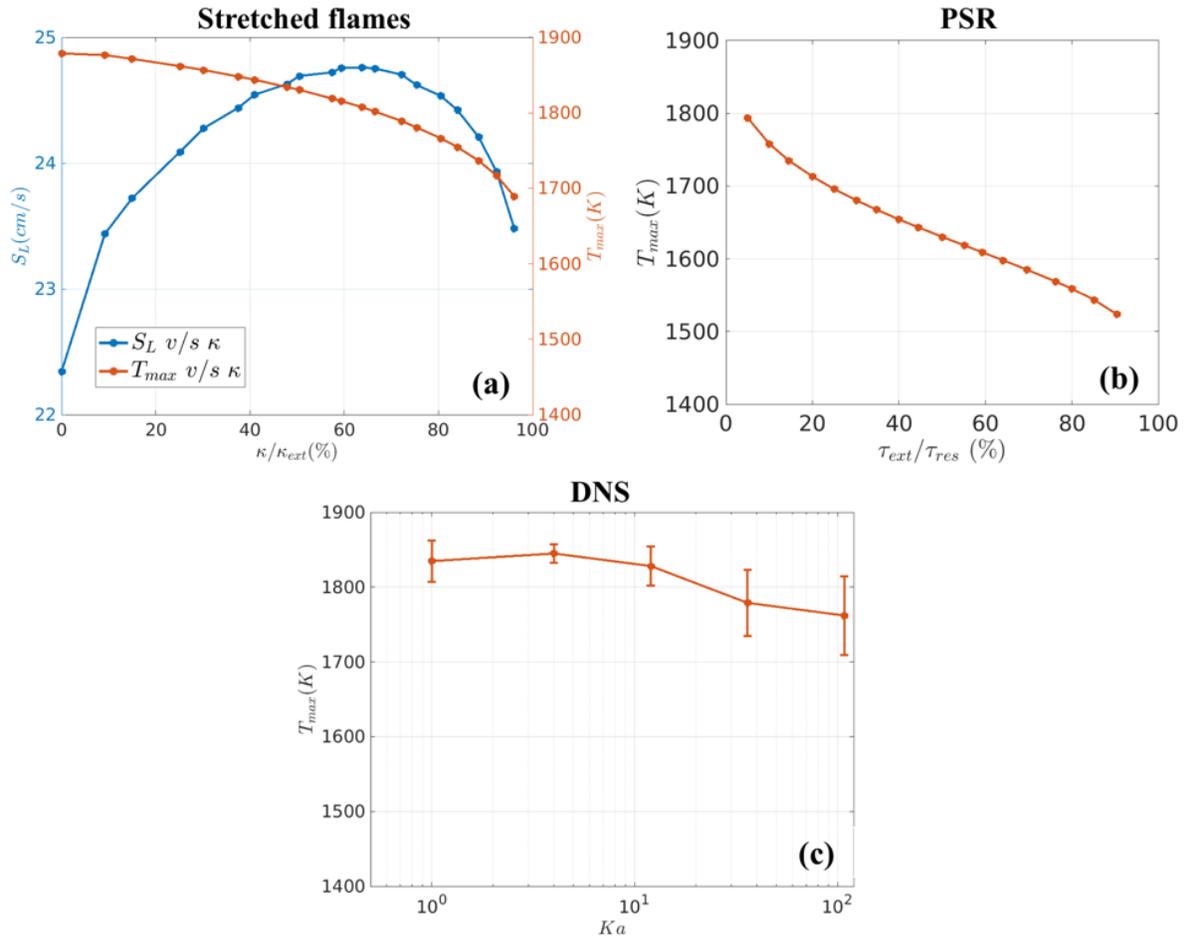

**Figure 2. (a) Flame speed and maximum temperature variation as a function of stretch rates, obtained from OPPDIF (b) maximum temperature variation as a function of residence time for PSR and (c) maximum temperature variation with increasing turbulence intensity for DNS calculation, $\phi$=0.7 n-dodecane/air, $T^u$=298K, $p$=1atm.**

## 3. Results and Discussions

In order to assess chemical pathways and the degree to which they are influenced by turbulence, we consider global (i.e., spatially integrated over the entire reaction zone) and local measures to quantify these changes.

### (A) Spatially Integrated comparisons of flame models and DNS

Figure 3 plots the variation of normalized heat release for stretched flames, the PSR and DNS results. Starting with Figure 3(a) and (b), the first observation is that the same set of dominant heat release reactions appear for both the stretched flame and PSR reference calculations. However, there are differences in sensitivity to $\kappa$ and $\tau_{res}$, as well as in the dominant reaction at a given point, between the two reference calculations. For example, the reaction $HO_2+OH \rightarrow O_2+H_2O$ increases with $\kappa$ with a maximum variation of ~8%. The same reaction decreases with decreasing residence time by ~65% for the PSR. Similarly, the reaction $CH_3+O \rightarrow CH_2O+H$ is fairly invariant stretch, changing by ~1% from $\kappa/\kappa_{ext}=0$ to $\kappa/\kappa_{ext} = 0.97$. The contribution of the same reaction increases by ~50% with decreasing PSR residence time. This observation is very different from that observed in the case of lighter fuels, such as hydrogen[2] and methane[9], wherein the variation between the contribution of a given reaction for stretched flame and for PSR was within ~20%.

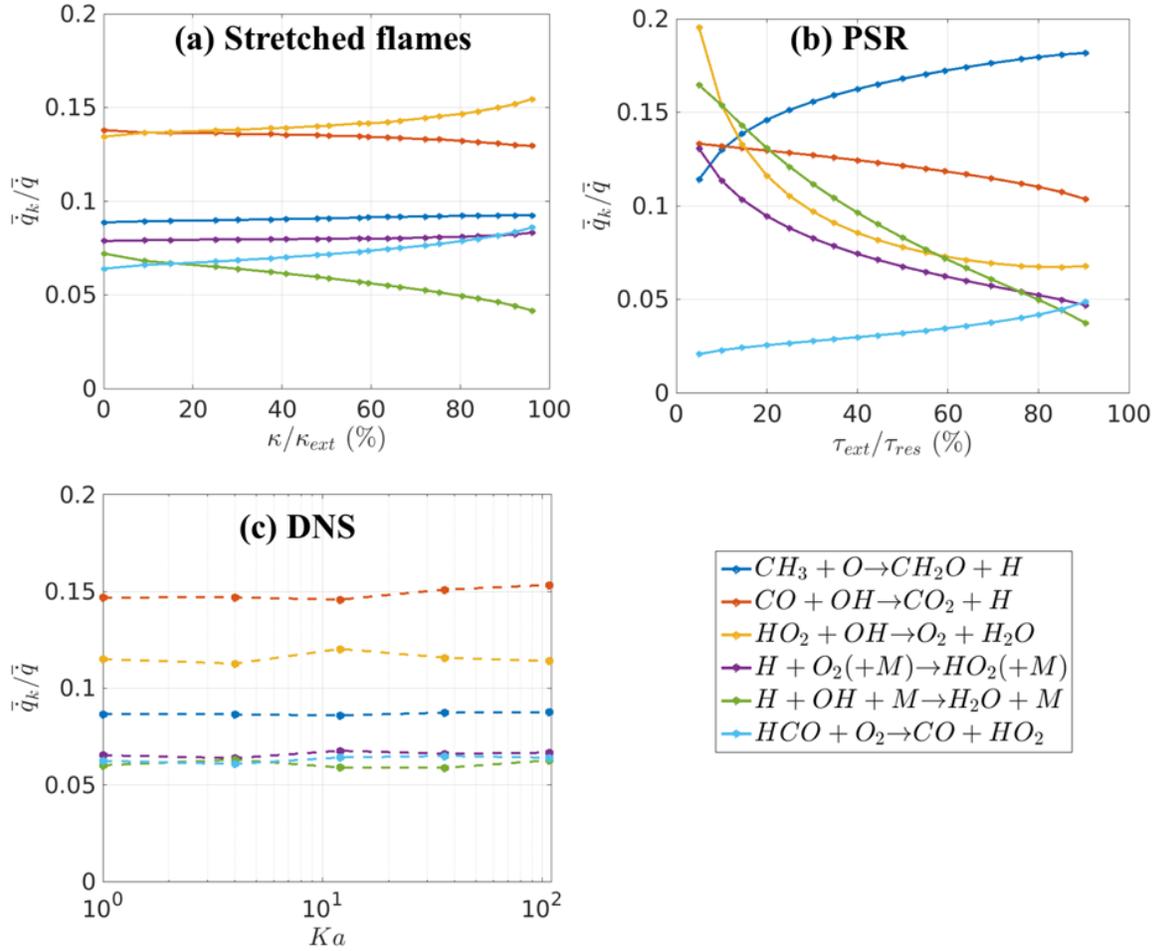

**Figure 3. Variation of normalized heat release with (a) increasing stretch and (b) decreasing residence time and (c) increasing turbulence intensities using You et al.[15]'s mechanism.**

Considering next the turbulent flame results in Figure 3(c), the plot shows that the dominant heat release reactions here are the same as those identified for stretched flames and PSR. Additionally, the ordering and behavior of the reactions closely resemble those of stretched flames. Overall, however, the DNS results show almost no change of the reactions' contribution with $Ka$. For example, the variation in normalized heat release for the dominant heat release reaction, $CO+OH \rightarrow CO_2+H$, with turbulence intensity is ~6% between $Ka = 1$ and $Ka = 12$. The difference between the unstretched, laminar case and the $Ka$=108 case is ~1%. Similarly, the maximum variation for the second dominant heat release reaction, $HO_2+OH \rightarrow O_2+H_2O$, is ~5% between $Ka$=1 and $Ka$=12.

Figure 4 plots the fractional consumption of the high temperature radicals, $CH_3$ (right) and HCO (left), for the DNS and model reactor/flame calculations. Unlike, the heat release results, significant quantitative variation of consumption rates via these reactions is observed between the stretched flame and PSR configurations. For stretched flames, $HCO+O_2 \rightarrow CO+HO_2$ surpasses $HCO+M \rightarrow CO+H+M$ as the dominant HCO-consuming reaction at high stretch rates. These two reactions are also the dominant HCO consumers for perfectly stirred reactors and have similar directional sensitivity to stretch/residence time for both models. For example, HCO consumption via the reaction, $HCO+M \rightarrow CO+H+M$, decreases by ~8% from $\kappa/\kappa_{ext}=0$ to $\kappa/\kappa_{ext} = 0.97$ and by ~32% from $\tau_{ext}/\tau_{res}=0.05$ to $\tau_{ext}/\tau_{res} = 0.91$. Similarly, the normalized rate of HCO consumption by $HCO+O_2 \rightarrow CO+HO_2$ increases by ~20% from $\kappa/\kappa_{ext}=0$ to $\kappa/\kappa_{ext} = 0.97$ and by ~65% from $\tau_{ext}/\tau_{res}=0.05$ to $\tau_{ext}/\tau_{res} = 0.91$.

The two dominant $CH_3$-consuming reactions are $CH_3+O \rightarrow CH_2O+H$ and $CH_3+OH \rightarrow CH_2^*+H_2O$. These two reactions account for almost ~65% and ~90% of the consumption of $CH_3$ for stretched flames and the perfectly stirred reactor, respectively. For the stretched laminar flame, both reactions show limited sensitivity to increasing stretch, with a maximum change of 4%. For the PSR, this change is ~25% with decreasing residence times.

The dominant reactions for each species in the turbulent flames exhibit a strong qualitative similarity with their stretched flame counterparts. For example, in Figure 4(a), $HCO+M \rightarrow CO+H+M$ consumes ~45% of the total HCO and changes by <1% with increasing turbulence intensity. The maximum change of ~8% is seen for the reaction $HCO+O_2 \rightarrow CO+HO_2$ between $Ka$=1 and $Ka$=12. Unlike the stretched laminar flames, no cross-over is seen by the two dominant reactions. In Figure 4(b), the dominant $CH_3$ consuming reaction of $CH_3+O \rightarrow CH_2O+H$ shows a maximum increase of ~7% with increasing turbulence intensity. The primary difference between stretched flames and DNS is the appearance of the fuel fragments recombination reactions of $2CH_3 \rightarrow C_2H_5+H$ and $CH_3+C_2H_4 \rightarrow nC_3H_7$ for the turbulent flames in addition to $2CH_3(+M) \rightarrow C_2H_6(+M)$.

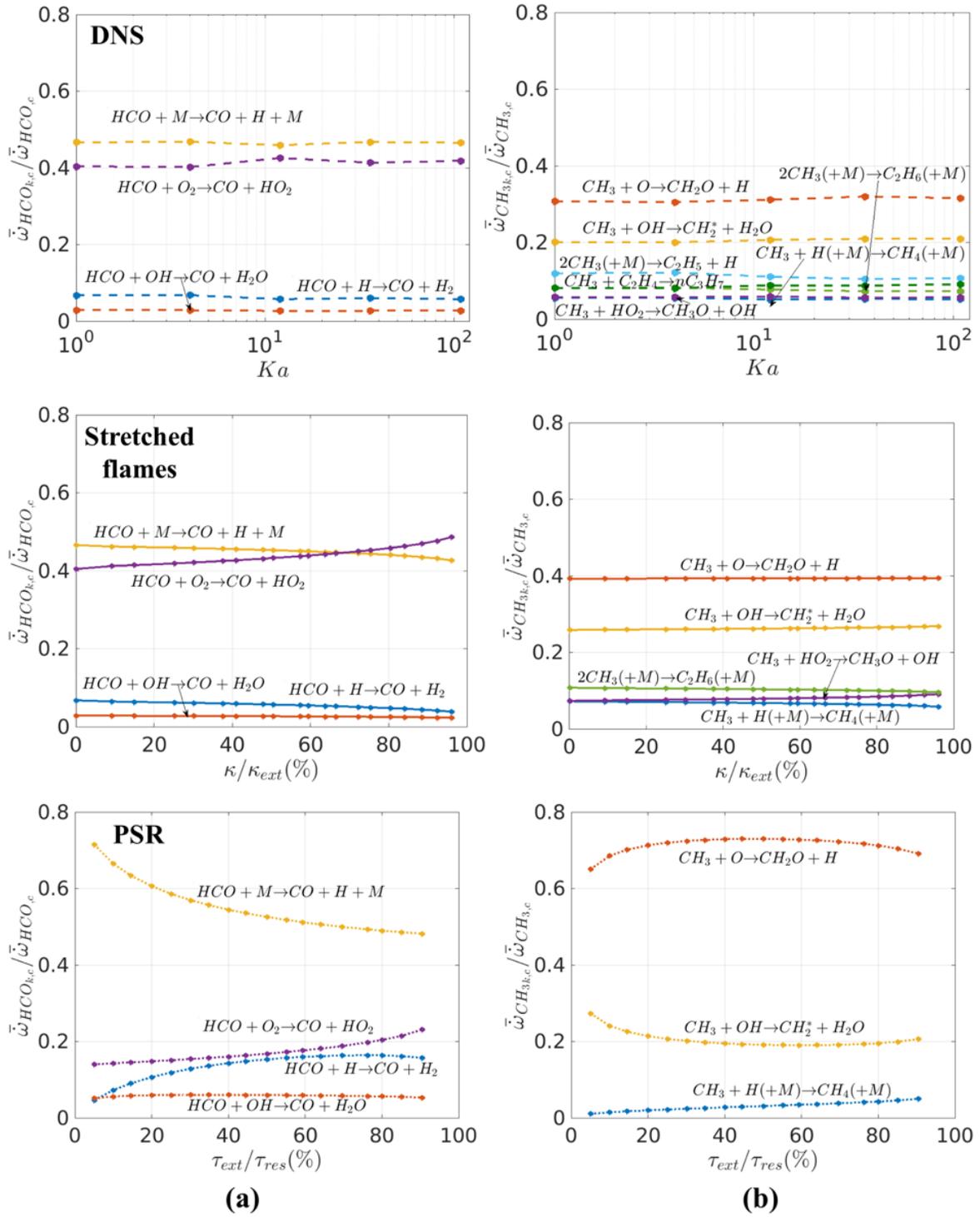

**Figure 4.** Dependence of normalized consumption rates for (a) HCO and (b) CH$_3$ upon Karlovitz number (left), stretch rate (center) and residence time (bottom).

Figure 5 plots the consumption rates for OH, a high temperature radical and HO₂ a, low temperature radical. Similar conclusions as discussed above can be drawn here- most notably, the weak effect of *Ka* on the relative contributions of radical formation/destruction reactions.

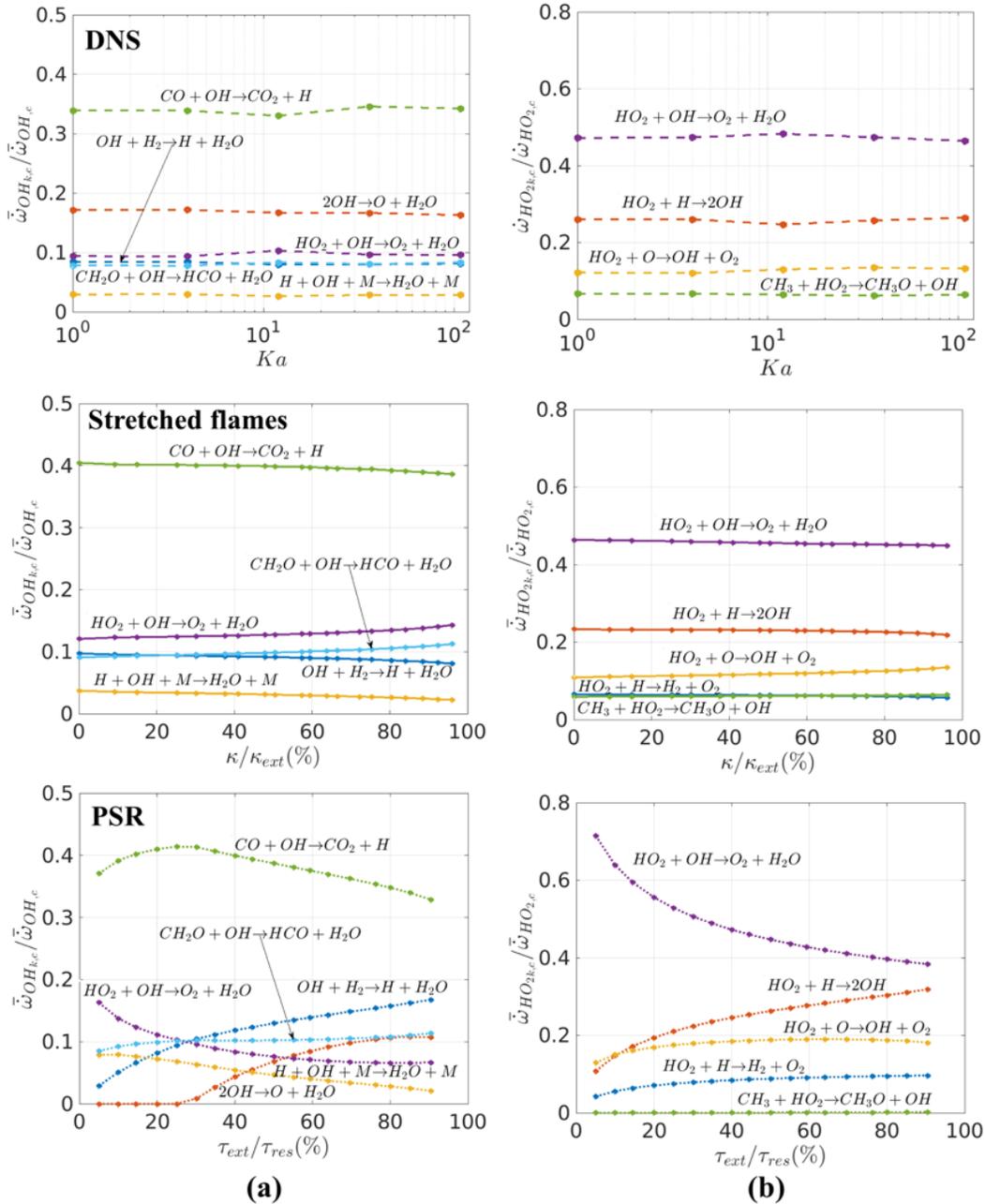

**Figure 5. Variation of normalized consumption rates for (a) OH and (b) HO₂ with increasing turbulence intensities (top), increasing stretch(center) and decreasing residence time(bottom).**

In general, all reactions for the turbulent and the stretched laminar flames show only minor variations with increasing turbulence intensity/ stretch. On the other hand, substantial variation is seen with changing residence times for the PSR. This is likely a thermal effect, as Figure 2 shows a significantly larger change in temperature with PSR residence time than for the stretched, laminar flame and the turbulent flames.

**(B)    Topologically conditioned results**

This section focuses on curvature conditioning of the integrated metrics discussed above. Five different topological regions (concave/convex spherical elements, concave/convex cylindrical elements, and saddle-points) conditioned on the two principal components of curvature, $\Bbbk_1$ and $\Bbbk_2$ ($\Bbbk_1 > \Bbbk_2$) are defined, as shown in Figure 6. The arrows indicate the direction of "flame" propagation.

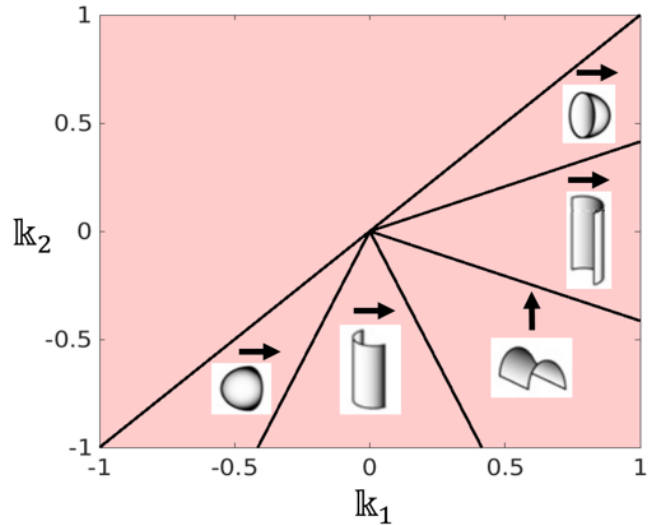

**Figure 6. Illustration of potential curvature based topologies.**

Figure 7 below plots the variation of fractional contribution of each of these five elements to the net heat release.

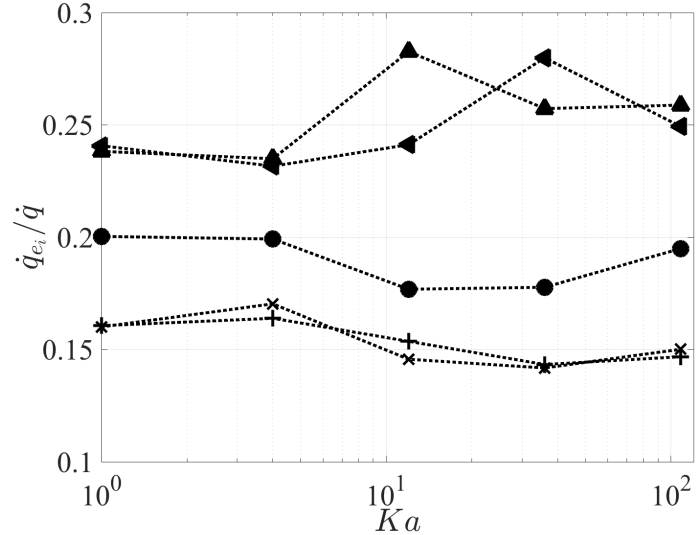

**Figure 7. Variation of fractional heat release within each element with Karlovitz number. (.▲:Spherical negatively-curved, .◄:Cylindrical negatively-curved, .●:Saddle-points, +:Spherical positively-curved, .x: Cylindrical positively-curved)**

It can be observed that most of the heat release occurs in the negatively-curved regions, as expected for this reactant mixture, where $Le>1$ and where burning is enhanced in negatively-curved regions. For $Le>1$, the mixture thermal diffusivity is higher than the mass diffusivity of the deficient fuel. As a result, in negatively-curved regions, the reactants lose species to the flame less rapidly compared to heat gained from the flame, resulting in stronger burning in these regions. The opposite behavior is seen for $Le>1$ positively-curved element.

Figure 8 plots the fractional heat release contribution of various reactions in each element. The heat release by each reaction is normalized by the total heat release within an element. It can be seen that the same reactions dominate the heat release in all 5 elements. In the positively-curved elements, the dominant reaction changes from $CO+OH \rightarrow CO_2+H$ to $HO_2+OH \rightarrow O_2+H_2O$ with increasing turbulence intensity. The most significant change is seen for the reaction $H+OH+M \rightarrow H_2O+M$ with a decrease in contribution of ~50% in the positively-curved elements. $H+O_2(+M) \rightarrow HO_2(+M)$ and $HCO+O_2 \rightarrow CO+HO_2$ change by ~40% with increasing turbulence intensity in these elements. Additionally, note the growing contribution of the reaction $HCO+O_2 \rightarrow CO+HO_2$ in these elements. This reaction has a higher heat release contribution than H+O$_2$(+M)→HO$_2$(+M) and H+OH+M→H$_2$O+M at higher turbulence intensities, suggesting a slight alteration of the heat release pathway at higher *Ka* for the positively-curved elements.

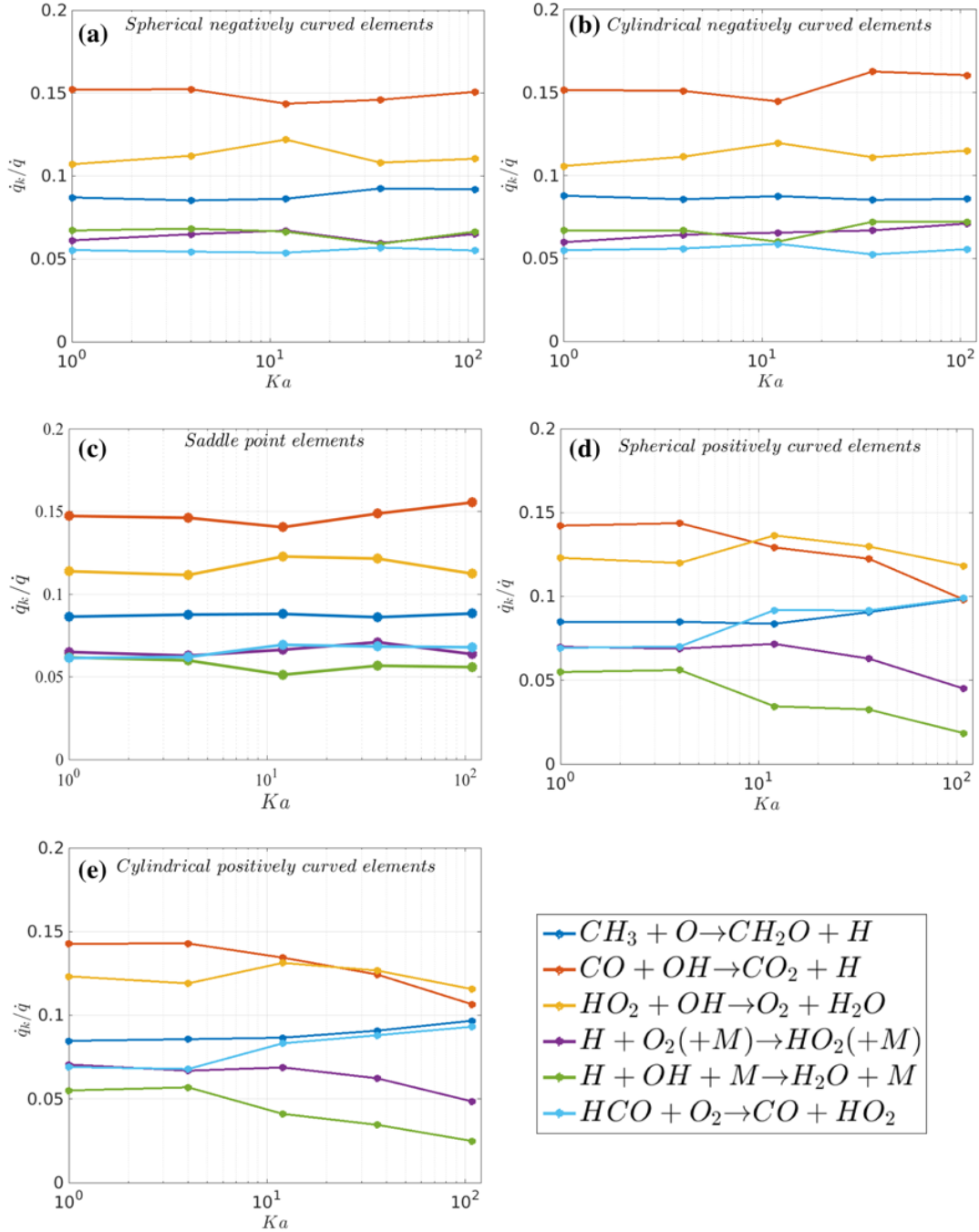

**Figure 8. Normalized heat release by the dominant reactions in (a) Spherical negatively-curved elements (b) Cylindrical negatively-curved elements (c) Saddle-point elements (d) Spherical positively-curved elements (e) Cylindrical positively-curved elements.**

The contribution of the reactions in the negatively-curved and saddle-point elements behave similar to their global counter-part. For example, the contribution of the reaction $CO+OH \rightarrow CO_2+H$ changes by ~6% for the global characteristics and for these three elements.

Figure 9 plots the normalized rate of consumption/production of certain key species. The same dominant reactions for HCO can be observed in Figure 9(a). For HCO consumption, note that $HCO+O_2 \rightarrow CO+HO_2$ is the dominant HCO consuming reaction in the positively-curved elements (Figure 9(a, left)) followed by $HCO+M \rightarrow H+CO+M$. The contribution of $HCO+O_2 \rightarrow CO+HO_2$ increases by ~25% with increasing turbulence intensities, whereas the contribution of $HCO+M \rightarrow H+CO+M$ decreased by ~25%. The same two reactions are observed for the negatively-curved elements and saddle-point elements (Figure 9(a, right)). The order of the dominant reactions is reversed for these elements and show limited sensitivity to turbulence intensity. For OH we observe modifications in the contributions of the secondary reactions with increasing turbulence intensity for the positively-curved elements (Figure 9(b, left)). The consumption of OH by $CO+OH \rightarrow CO_2+H$ decreases by ~15% with increasing turbulence intensity. At higher turbulence intensities, the reactions $HO_2+OH \rightarrow H_2O+O_2$, $OH+H_2 \rightarrow H_2O+H$ and $CH_2O+OH \rightarrow HCO+H_2O$ have a higher consumption rate than $2OH \rightarrow H_2O+O$, which is the second dominant OH consumer at lower turbulence intensities. The contributions of these reactions do not change significantly for the negatively-curved and saddle-point elements (Figure 9(b, right)). The pathways for water formation, one of the key products in hydrocarbon combustion, are strongly affected with increasing turbulence intensity for the positively-curved elements (Figure 9(c, left)). For example, $CH_2O+OH \rightarrow HCO+H_2O$ takes over as the dominant $H_2O$ producing reaction from $HO_2+OH \rightarrow H_2O+O_2$ at higher turbulence intensity. Also, the reactions $C_2H_4+OH \rightarrow C_2H_3+H_2O$ and $CH_3+OH \rightarrow CH_2^*+H_2O$ have a higher contribution to $H_2O$ production than $2OH \rightarrow H_2O+O$ with increasing $Ka$. These contributions are increased by ~25% from $Ka=1$ to $Ka=108$. Other species such as $HO_2$, $CH_3$ show limited sensitivity to turbulence and curvature.

To summarize, the largest variations in reaction pathways with increasing turbulence intensity occur for positively-curved elements; variations for the negatively-curved elements and saddle-point elements are much weaker.

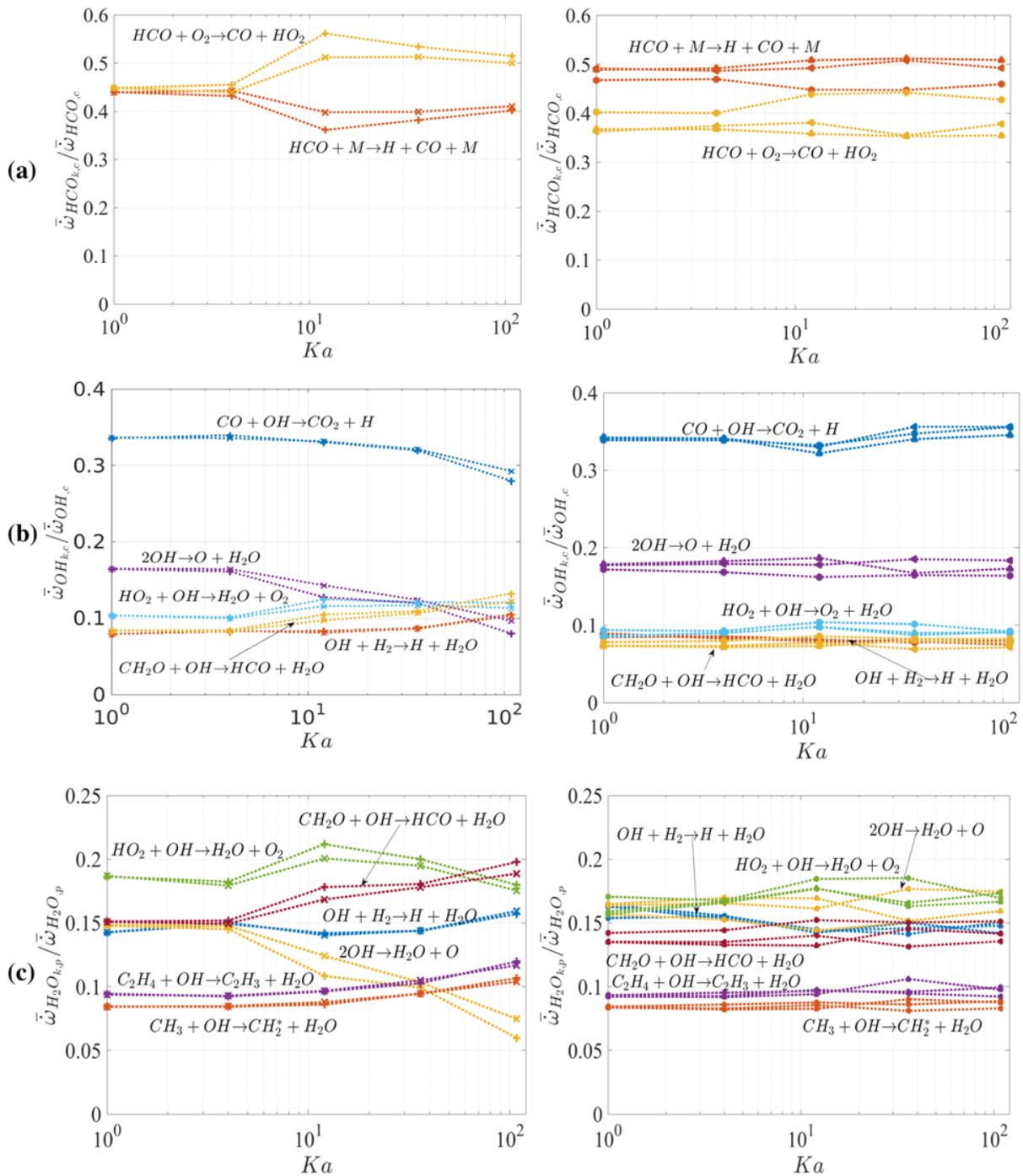

**Figure 9.** Normalized species consumption by the dominant reactions for (a) HCO consumption (b) OH consumption (c) $H_2O$ productions. (.▲:Spherical negatively-curved, .◄:Cylindrical negatively-curved, .●:Saddle-points, +:Spherical positively-curved, .x: Cylindrical positively-curved).

## (C) Local Analysis: Temperature conditioned heat release and reaction rates

Figure 10 shows spatial profiles of temperature and two of the reaction rates. The clear thickening of the thermal extent of the flame with increasing turbulence intensity, as established in literature [10-12], is evident. For the lower *Ka* cases, the flame is thin and slightly wrinkled. With increased turbulence intensities, the spatial broadening of the temperature region between 800-1300K can be observed. On the other hand, not all reaction layers thicken appreciably. For example, the figure shows that the reaction $HO_2+OH \rightarrow O_2+H_2O$ occurs within a thin reaction zone over the entire *Ka* range. However, profiles of some reactions tend to broaden along with the temperature field. This can be seen, for example in reaction $CO+OH \rightarrow CO_2+H$ which has a broadened reaction zone with increasing turbulence intensity and shows a pronounced effect at *Ka*=108.

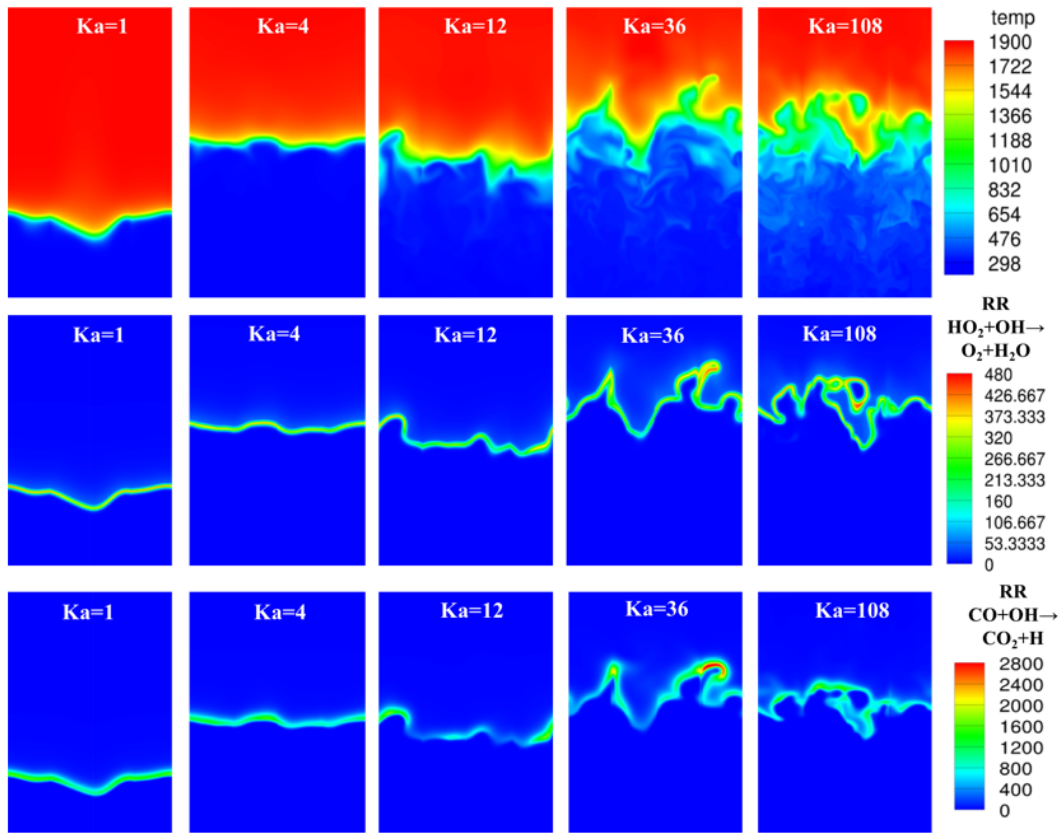

**Figure 10. Slices of temperature(top), $HO_2+OH \rightarrow O_2+H_2O$ reaction rate (center) and $CO+OH \rightarrow CO_2+H$ reaction rate(bottom). The slices are constructed using the x=0 and y=0 plane using the fact of periodic lateral boundary conditions.**

In Section (A), we showed that the integrated variation of the reaction rates and heat release by various reactions exhibit limited sensitivity to increasing turbulence intensity. We thus, examine the reaction rate profiles in temperature space, and compare them with two reference cases - the unstretched laminar, and the maximally stretched laminar (i.e. $\kappa/\kappa_{ext} \approx 0.97$) cases. To do this, we divide the temperature into bins of 25K from 298K to 1898K and plot the average values in each bin.

Figure 11 plots the net heat release and fuel consumption as a function of temperature for *Ka*=1,12 and 108 (only three cases are plotted for clarity). For reference, the unstretched and highly stretched laminar flame results are plotted.

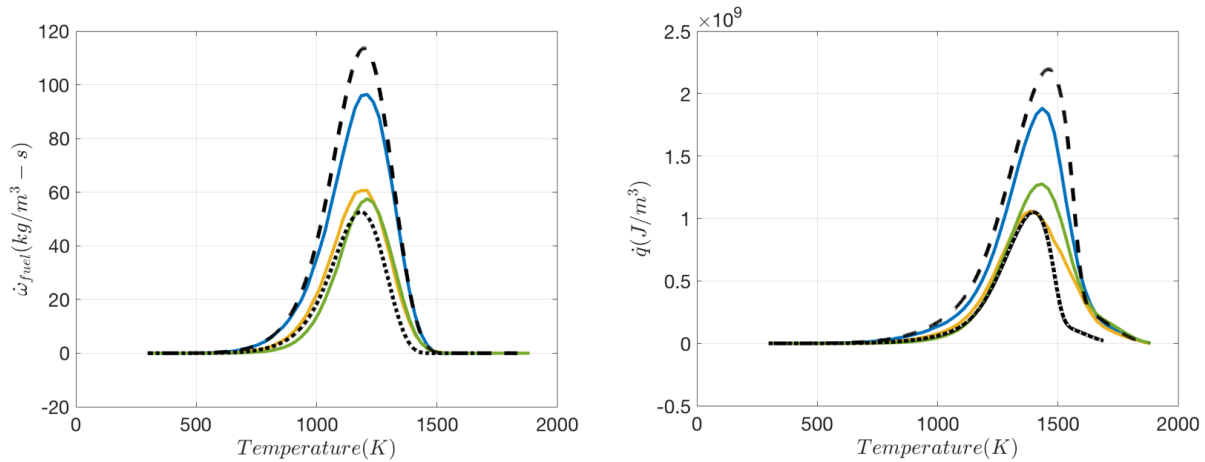

**Figure 11. Variation of fuel consumption (left) and heat release (right) with temperature. *Ka*=1(Solid blue line), *Ka*=12(Solid yellow line), *Ka*=108 (Solid green line), Unstretched laminar flame (Dashed black line), Highly stretched laminar flame (Dotted black line).**

The figure indicates that the peak temperature for fuel consumption and heat release do not significantly change with turbulence intensity. However, there is a slight shift in heat release towards high temperature. For heat release, the turbulent flame profiles are well-represented by the extreme laminar cases, as seen in Figure 11(right). As expected, the *Ka*=1,4 cases are closer in behavior to the unstretched laminar flame profile. The higher turbulence cases (*Ka*=12-108) behave similar to the highly stretched laminar flame up to a temperature of ~1200K. The highly stretched laminar flame reaches a lower equilibrium temperature than the unstretched/turbulent cases, resulting in its deviation from the *Ka*=12-108 results at temperatures beyond 1400K.

Figure 12 below plots the variation of reaction rates of some of the key reactions identified in the previous section.

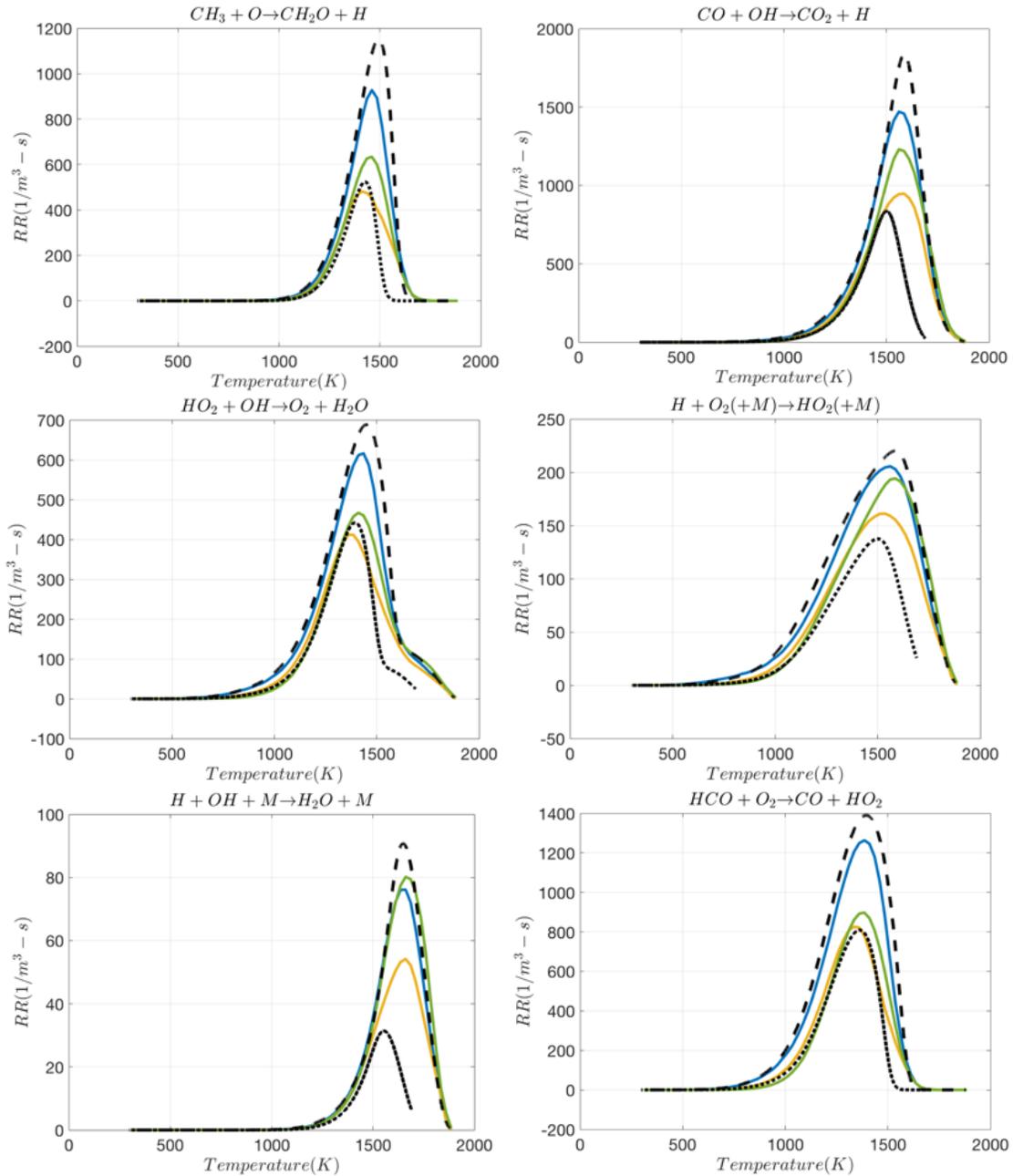

**Figure 12. Variation of reaction rates of different reactions with temperature.** *Ka*=1(Solid blue line), *Ka*=12(Solid yellow line), *Ka*=108 (Solid green line), Unstretched laminar flame (Dashed black line), Highly stretched laminar flame (Dotted black line).

The lower turbulence intensities of *Ka*=1 (and 4) follow the unstretched laminar profile well. With increasing turbulence intensities, however the trend is not obvious. Response of certain reactions such as HCO+$O_2$→CO+$HO_2$, $HO_2$+OH→$O_2$+$H_2O$, $CH_3$+O→$CH_2O$+H is similar to the highly stretched laminar flame whereas, the other reactions H+OH+M→$H_2O$+M, H+$O_2$(+M)→$HO_2$(+M), CO+OH→$CO_2$+H have a qualitatively similar response to the unstretched laminar flame (peak temperature, slopes). A non-monotonic change in the peak reaction rates with increasing turbulence intensity can be observed. For example, between *Ka*=1 and *Ka*=12 there is a reduction in peak reaction rate whereas between *Ka*=12 and *Ka*=108 an increase in the peak reaction rate can be noted. In general, we observe a shift of reaction rate profiles towards higher temperatures. This is more clearly illustrated in Figure 13.

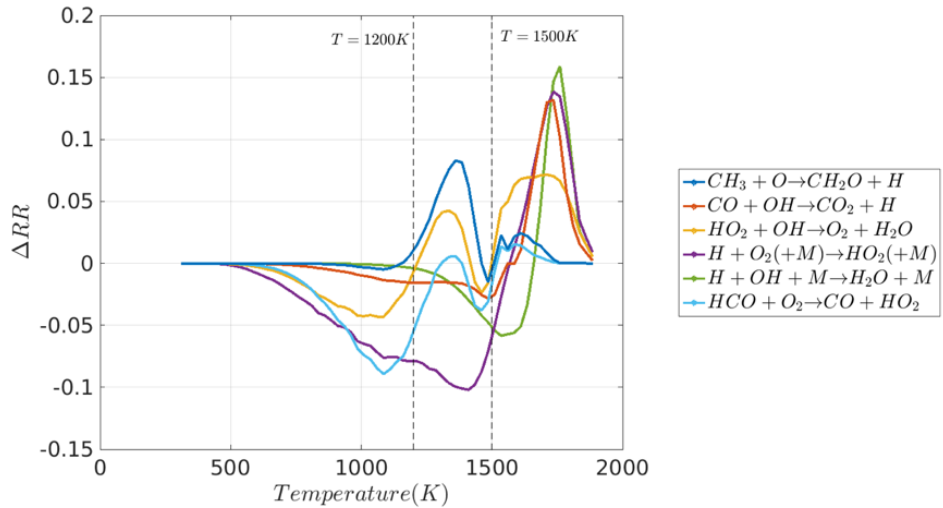

**Figure 13. Variation of change of reaction between *Ka*=1 and *Ka*=108 for different reactions.**

Figure 13 plots the change in reaction rates($\Delta RR$) between the two extreme turbulent cases; i.e. *Ka*=1 and *Ka*=108. $\Delta RR$ is calculated by subtracting the normalized reaction rate profile of *Ka*=1 from the normalized reaction rate profile of *Ka*=108. This shows a positional shift of the profiles in temperature space i.e. if the profiles shift with increasing turbulence intensity. Thus, $\Delta RR < 0$ indicates higher normalized reaction rates for *Ka*=1 than for *Ka*=108. For reference, T=1200K and T=1500K iso-lines are indicated. The reactions that are active below 1200K, such as HCO+$O_2$→CO+$HO_2$, $HO_2$+OH→$O_2$+$H_2O$, and H+$O_2$(+M)→$HO_2$(+M), show reduced activity in this temperature range at higher turbulence intensities. These same reactions have a stronger high temperature sensitivity i.e. $\Delta RR > 0$, indicating increased activity in this region. On the other hand, the reactions which are active above 1200K, show very little deviation below 1200K. For example, the high-temperature reaction, $CO+OH \rightarrow CO_2+H$, shows much higher rates above 1500K. This behavior is also seen for the other high-temperature reaction, $H+OH+M \rightarrow H_2O+M$. In fact, this latter reaction shows lower rates for much of the temperature range below 1600K. The reaction, $CH_3+O \rightarrow CH_2O+H$, exhibits an increased activity in the interim range, between 1200-1500K. It also shows increased rates above 1500K (though not as strong as in the interim region).

An interesting feature is observed for reactions involving fuel fragments whose rates peak in the low-temperature region (i.e. below 1200K). Figure 14 plots the variation of reaction rates for two representative reactions involving fuel fragments, $pC_4H_9$ and $nC_3H_7$.

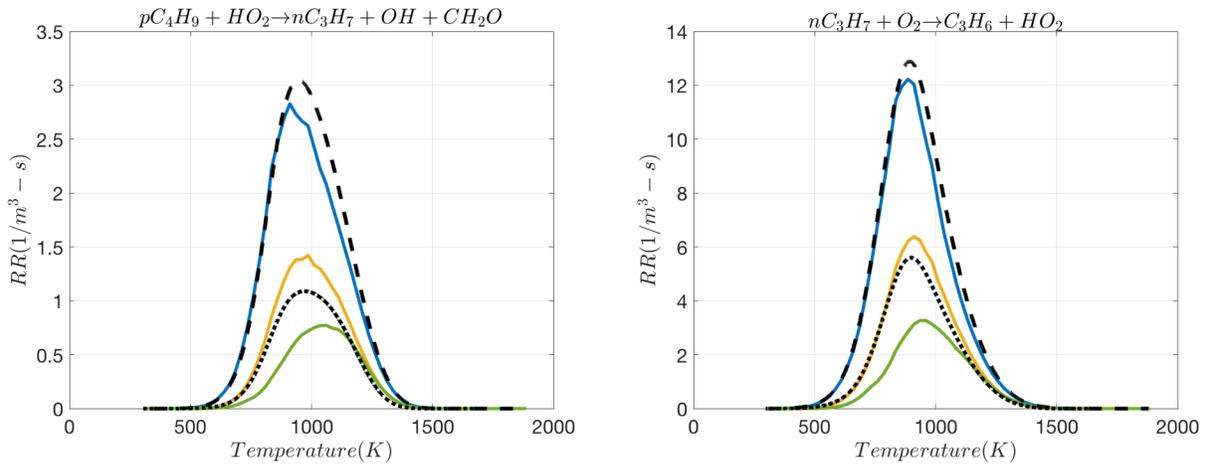

**Figure 14. Variation of reaction rates with temperature. *Ka*=1(Solid blue line), *Ka*=12(Solid yellow line), *Ka*=108 (Solid green line), Unstretched laminar flame (Dashed black line), Highly stretched laminar flame (Dotted black line).**

A clear systematic shift of the reaction rate profiles towards higher temperature region with increasing turbulence intensity can be observed in Figure 14. This behavior is consistent for all the reactions whose rates peak below 1200K (the temperature of peak fuel consumption).

Overall, the response of the reactions to increased turbulence intensities is well represented by the two extreme cases of laminar stretched flames, namely the unstretched laminar case and the most stretched case ($\kappa/\kappa_{ext} \approx 0.97$) when plotted in temperature space. With increasing turbulence intensities all reactions show a movement towards higher temperature. This shift maybe related to the species concentrations which are altered by turbulence resulting in changes in the reaction rates which depend on the cross correlation between the reactant species concentrations. Figure 15 plots the conditional concentration means for $nC_3H_7$(fuel fragment), $HO_2$(low temperature radical), CO (high temperature stable species) and OH (high temperature species).

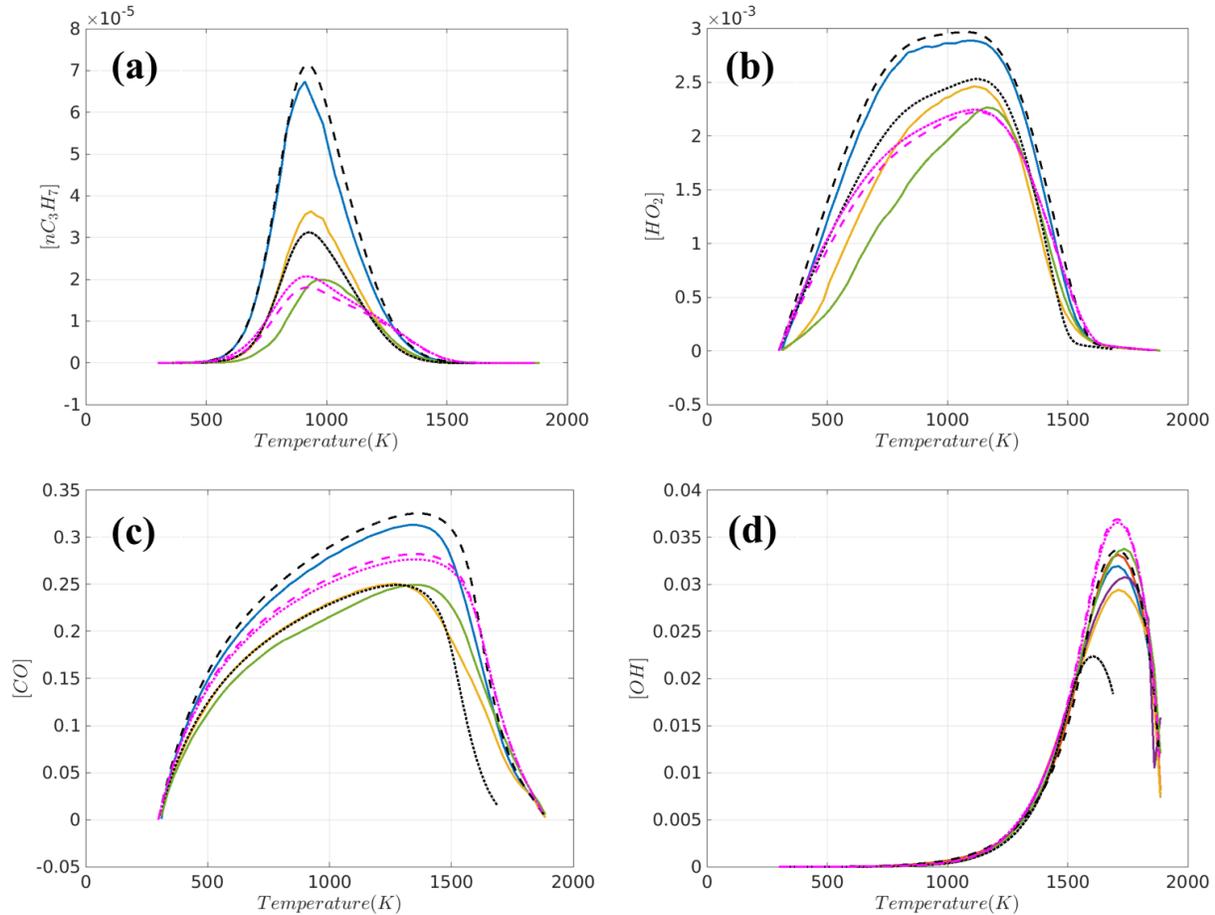

**Figure 15. Variation of concentration for (a) $nC_3H_7$ (b) $HO_2$ (c) CO (d) OH with temperature. *Ka*=1(Solid blue line), *Ka*=12(Solid yellow line), *Ka*=108 (Solid green line), Unstretched laminar flame (Dashed black line), Highly stretched laminar flame (Dotted black line), Unstretched laminar flame with Le=1 (Dashed magenta line), Highly stretched laminar flame with Le=1 (Dotted magenta line).**

It can be observed in Figure 15 that the concentration profiles of $nC_3H_7$, $HO_2$ and CO shift towards higher temperatures with increased turbulence intensity and this directly influences the behavior of the reactions involving these species. For example, for the reaction $nC_3H_7+O_2 \rightarrow C_3H_6+HO_2$, since $O_2$ concentration profile (not shown here) has a response similar to OH and shows limited sensitivity to turbulence, a direct correlation can be seen between the reaction rate profile in Figure 14 and the species concentration profile for $nC_3H_7$ in Figure 15. The shift of the species profile can be partly understood by comparing the profiles with laminar calculations using $Le$=1 transport. For the species presented, the $Le$=1 profiles shift towards higher temperatures compared to their laminar mixture-averaged counterpart. Even though the $Le$=1 profiles do not replicate the behavior of the turbulent flames (at higher turbulence intensities), they provide a first indication of the effect of increased diffusivity due to turbulence on the chemical structure of the flame. This idea will be explored in our future work.

## 4. Conclusions

This paper examines the effects of turbulence on the chemical pathways for lean premixed $n$-dodecane/air turbulent flames. It is observed that the fractional contribution of the dominant heat release reactions changes very little with increasing turbulence intensity, even at turbulence levels where the flame structure is significantly disrupted. For example, the reaction, $CO+OH \rightarrow CO_2+H$ which accounts for ~15% of the total heat release shows limited variation (~6%) for $Ka$ varying from 0 to 108. The $H_2O$ formation reaction, $HO_2+OH \rightarrow O_2+H_2O$, contributes about ~12% to the total heat release, and increases by ~5% over the same range.

For the curvature-conditioned results, it is observed that the negatively-curved and the saddle-point elements behave similar to their global counterpart. For example, the contribution of the reaction $CO+OH \rightarrow CO_2+H$, which accounts for ~15% of the total heat release, changes by ~6% for the global characteristics and for these three elements. However, more significant changes are observed for the positively-curved elements. The most dominant heat release reaction changes from $CO+OH \rightarrow CO_2+H$ to $HO_2+OH \rightarrow O_2+H_2O$ in the positively-curved elements with increasing turbulence intensities. The most significant change is seen for the reaction $H+OH+M \rightarrow H_2O+M$, with a decreased contribution of ~50% in the positively-curved elements. A stronger sensitivity of the reaction path for different species is observed in these elements as well. For example, the contribution of the second dominant OH consuming reaction, $2OH \rightarrow H_2O+O$, decreases by ~50% with increasing turbulence intensity, whereas in the negatively-curved elements it changes by ~5%. Certain species show altered pathway between the different elements. For example, the dominant HCO consuming reaction for the positively-curved elements is HCO+$O_2$→CO+$HO_2$. The dominant reaction changes to HCO+M→H+CO+M in the saddle-point and negatively-curved elements.

The turbulent flame results mirror closely those of the stretched flames. For example, the dominant $CH_3$-consuming reaction, $CH_3$+O→$CH_2$O+H, is responsible for 65% of the total $CH_3$ consumption, and changes by ~7% with increasing turbulence intensity. The same reaction changes by ~4% from $\kappa/\kappa_{ext}$=0 to $\kappa/\kappa_{ext} = 0.97$ in the steady stretched flames.

Overall, turbulence does not seem to affect the integrated reaction rates or heat release. However, local analysis reveals effects that are not reflected in these integrated quantities. All reactions have increased reaction rates at higher temperatures (i.e. >1200K) and reduced rates at lower temperatures. The most dramatic change is observed for reactions involving fuel fragments with peak rate near 900K.

Several questions remain for future work. A comparison of the effects of turbulence on different fuels is warranted. Earlier studies indicate a higher low temperature activity for light fuels such as $H_2$ and to some extent $CH_4$[4, 5]. The current study shows a different effect on *n*-dodecane flames. It would be interesting to look at a comparative local analysis for all these fuels. Additionally, further investigation is necessary to understand the non-monotonic change of the turbulent profiles with increasing turbulence intensity for *n*-dodecane. Finally, consequences of this understanding on chemistry models for simulations is required. The chemistry models heavily rely on laminar calculations for validations and it is essential to verify if a limited set of progress and controlling variables can capture all the important features of the flame.

## 5. Acknowledgement

The work was funded by the Air Force Office of Scientific Research (contract #FA9550-16-1-0442), contract monitor Dr. Chiping Li. This work used computational resources of charge# TG-CTS160017 under Project PI Dr. Vishal Acharya at the Extreme Science and Engineering Discovery Environment(XSEDE), which is supported by National Science Foundation grant number ACI-1548562[22].

**Appendix A: Sensitivity to reaction mechanisms**

An analysis of the effects of turbulence on details of lean premixed *n*-dodecane flames is based on comparisons between turbulent flame simulations (Aspden et al. [11]) and various low-dimensional and steady idealized configurations. Both the turbulent results and our subsequent computations for the idealized cases, were based on the detailed model of You et al.[15] for reaction kinetics, thermodynamic relationships and transport coefficients. Since there are a number of distinct models in the literature for this fuel in this regime, it is reasonable to ask whether the results of our study might be sensitive to which model was used. As an alternative to repeating the entire study, including the referenced costly DNS calculations, with each of the published models, we explore here a set of representative flames in the simplified configuration only, and focus the comparison on the reaction rate data that is key to the analysis above. Here, results of the You model are compared to those of two others, Luo et al.[16] and Narayanswamy et al.[17]).

Figure A1 and Figure A2 plot the fractional contribution of heat release for the three kinetic models for stretched flames and perfectly stirred reactors, respectively. These results reveal a varied ordering of the dominant heat release reactions. For example, in the case of You et al.[15], the reaction $HO_2+OH \rightarrow O_2+H_2O$ is the dominant heat release reaction for stretched flames. However, this reaction is the third dominant reaction for Luo et al.[16] and changes from third to second dominant reaction for Narayanswamy et al.[17] with increasing stretch rates. However, their behavior remained fairly consistent across the three mechanisms for the two laminar flame models of stretched flames and perfectly stirred reactors. For example, the fractional change in heat release with increasing stretch for the reaction $H+OH+M \rightarrow H_2O+M$ changes by ~40% for Luo et al.[16] , You et al.[15] and 35% for Narayanswamy et al.[17] These numbers for the reaction, $CO+OH \rightarrow CO_2+H$, are ~5% and ~6%, respectively. For perfectly stirred reactors, the reaction, $CH_3+O \rightarrow CH_2O+H$, is the dominant source of heat release for the L2 and Y1 mechanisms. However, this is the third (or second) dominant heat release reaction at higher (or lower) residence times for NS3. Again, there is a good qualitative match in the behavior of the reactions across the three mechanisms. For example, the reaction, $CH_3+O \rightarrow CH_2O+H$, shows a wide quantitative variation in its contribution to heat release. Its increased contribution to heat release however is fairly consistent across the three mechanisms varying from ~55% for Y1 and L2 to ~65% for NS3. The reaction, H+O$_2$(+M) →HO$_2$(+M), shows a smaller quantitative spread and its change in contribution is around ~60% for Y1 and L2 to ~70% for NS3. Thus, the analysis in this paper is likely insensitive to the choice of the detailed chemistry model used.

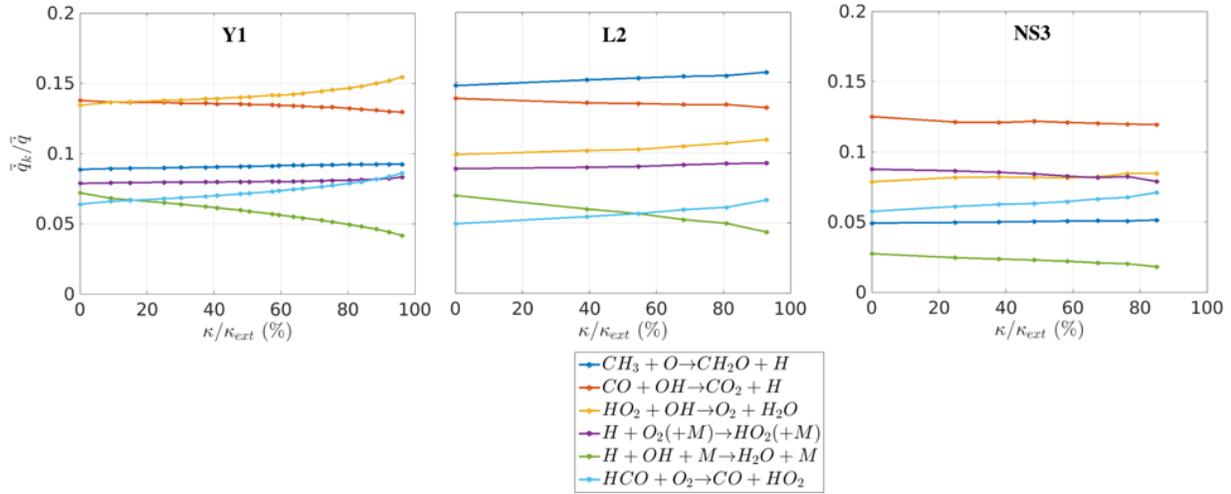

**Figure A1. Variation of normalized heat release with increasing stretch for three mechanisms. Y1: You et al., L2: Luo et al. NS3: Narayanswamy et al. All three plots have the same vertical scale.[23]**

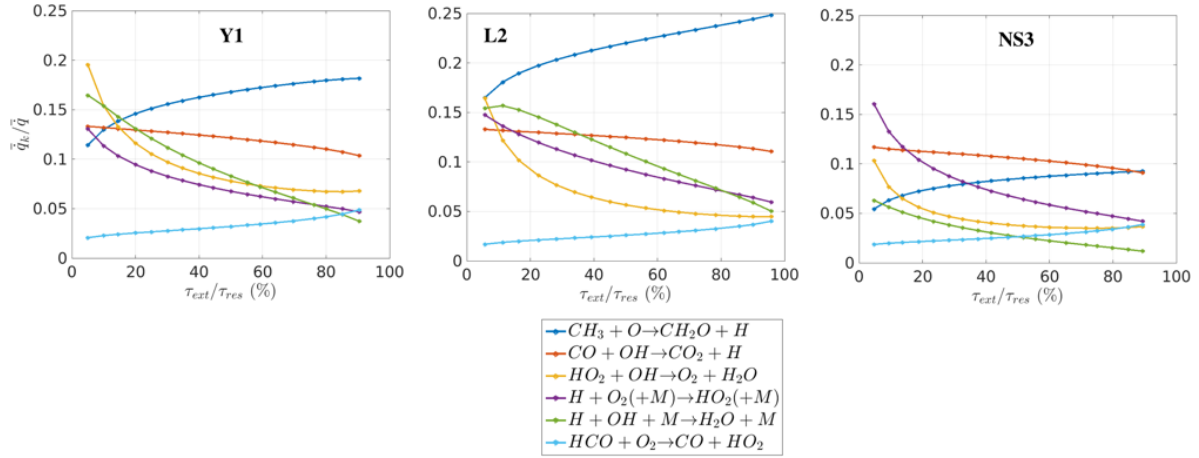

**Figure A2. Variation of normalized heat release with decreasing residence time for three mechanisms. Y1: You et al., L2: Luo et al. NS3: Narayanswamy et al. All three plots have the same vertical scale. [23]**

# 6. References

[1] J.A. van Oijen, A. Donini, R.J.M. Bastiaans, J.H.M. ten Thije Boonkkamp, L.P.H. de Goey, State-of-the-art in premixed combustion modeling using flamelet generated manifolds, Progress in Energy and Combustion Science 57 (2016) 30-74.

[2] D. Dasgupta, W. Sun, M. Day, T. Lieuwen, Effect of Turbulence-Chemistry interactions on chemical pathways for turbulent hydrogen-air premixed flames, Combustion and Flame 176 (2017) 191-201.

[3] M.S. Day, X. Gao, J.B. Bell, Properties of lean turbulent methane-air flames with significant hydrogen addition, Proceedings of the Combustion Institute 33 (2011) 1601-1608.

[4] A.J. Aspden, M.S. Day, J.B. Bell, Turbulence-chemistry interaction in lean premixed hydrogen combustion, Proceedings of the Combustion Institute 35 (2015) 1321-1329.

[5] H. Carlsson, R. Yu, X.-S. Bai, Direct numerical simulation of lean premixed CH4/air and H2/air flames at high Karlovitz numbers, International Journal of Hydrogen Energy 39 (2014) 20216-20232.

[6] D. Dasgupta, W. Sun, M. Day, T. Lieuwen, Investigation of chemical pathways for turbulent Hydrogen-Air premixed flames, 54th AIAA Aerospace Sciences Meeting, American Institute of Aeronautics and Astronautics, 2016.

[7] S. Lapointe, B. Savard, G. Blanquart, Differential diffusion effects, distributed burning, and local extinctions in high Karlovitz premixed flames, Combustion and Flame 162 (2015) 3341-3355.

[8] B. Savard, H. Wang, A. Teodorczyk, E.R. Hawkes, Low-temperature chemistry in n-heptane/air premixed turbulent flames, Combustion and Flame 196 (2018) 71-84.

[9] D. Dasgupta, W. Sun, M. Day, T. Lieuwen, Turbulence effects on the chemical pathways for premixed methane/air flames, 55th Aerospace Sciences Meeting, Grapevine, Texas, 2017.

[10] A.J. Aspden, M.S. Day, J.B. Bell, Three- dimensional direct numerical simulation of turbulent lean premixed methane combustion with detailed kinetics, Combustion and Flame 166 (2016) 266–283.

[11] A.J. Aspden, J.B. Bell, M.S. Day, F.N. Egolfopoulos, Turbulence-Flame Interactions in Lean Premixed Dodecane Flames, Proceedings of the Combustion Institute 36 (2017) 2005-2016.

[12] B. Savard, G. Blanquart, Broken reaction zone and differential diffusion effects in high Karlovitz n-C7H16 premixed flames, Combustion and Flame 162 (2015) 2020-2033.

[13] M.S. Day, J.B. Bell, Numerical simulation of laminar reacting flows with complex chemistry, Combustion Theory and Modelling 4 (2000) 535-556.

[14] J. Bell, M. Day, A. Almgren, M. Lijewski, C. Rendleman, R.Cheng, I. Shepherd, Simulation of Lean Premixed Turbulent Combustion, Journal of Physics: Conference Series 46 (2006) 1-15.

[15] X. You, F.N. Egolfopoulos, H. Wang, Detailed and simplified kinetic models of n-dodecane oxidation: The role of fuel cracking in aliphatic hydrocarbon combustion, Proceedings of the Combustion Institute 32 (2009) 403-410.

[16] Z. Luo, S. Som, S.M. Sarathy, M. Plomer, W.J. Pitz, D.E. Longman, T. Lu, Development and validation of an n-dodecane skeletal mechanism for spray combustion applications, Combustion Theory and Modelling 18 (2014) 187-203.

[17] K. Narayanswamy, P. Pepiot, H. Pitsch, A chemical mechanism for low to high temperature oxidation of n-dodecane as a component of transportation fuel surrogates, Combustion and Flame 161 (2014) 866-884.

[18] M. Day, J. Bell, P.T. Bremer, V. Pascucci, V. Beckner, M. Lijewski, Turbulence effects on cellular burning structures in lean premixed hydrogen flames, Combustion and Flame 156 (2009) 1035-1045.

[19] R.J. Kee, F.M. Rupley, J.A. Miller, M.E. Coltrin, J.F. Grcar, E. Meeks, H.K. Moffat, A.E. Lutz, G.D.-. Lewis, M.D. Smooke, J. Warnatz, G.H. Evans, R.S. Larson, R.E. Mitchell, L.R. Petzold, W.C. Reynolds, M. Caracotsios, W.E. Stewart, P. Glarborg, C. Wang, O. Adigun, CHEMKIN Collection, Release 3.6, Reaction Design, Inc., San Diego, CA, 2000.

[20] A.E. Lutz, R.J. Kee, J.F. Grcar, F.M. Rupley, OPPDIF: A Fortran Program for Computing Opposed-Flow Diffusion Flames, Sandia National Laboratories, Livermore, CA, 1996.

[21] P. Glarborg, R.J. Kee, J.F. Grear, J.A. Miller, PSR: A FORTRAN Program for Modelling Well-Stirred Reactors, Sandia National Laboratories, Livermore, CA, 1986.

[22] J. Towns, T. Cockerill, M. Dahan, I. Foster, K. Gaither, A. Grimshaw, V. Hazlewood, S. Lathrop, D. Lifka, G.D. Peterson, R. Roskies, J. Ray Scott, N. Wilkins-Diehr, XSEDE: Accelerating Scientific Discovery, Computing in Science and Engineering 16 (2014) 62-74.

[23] D. Dasgupta, W. Sun, M. Day, T. Lieuwen, Sensitivity of chemical pathways to reaction mechanisms for n-dodecane, 10th U. S. National Combustion Meeting, College Park, Maryland, 2017.